# Efficiency at maximum power of thermochemical engines with near-independent particles

Xiaoguang Luo,[1] Nian Liu,[2] and Teng Qiu[1*]

[1]*Department of Physics, Southeast University, 211189 Nanjing, China*

[2]*Department of Physical and Electronics, Anhui Science and Technology University, Bengbu 233100, China*

Two-reservoir thermochemical engines are established in by using near-independent particles (including Maxwell-Boltzmann, Fermi-Dirac, and Bose-Einstein particles) as the working substance. Particle and heat fluxes can be formed based on the temperature and chemical potential gradients between two different reservoirs. A rectangular-type energy filter with width $\Gamma$ is introduced for each engine to weaken the coupling between the particle and heat fluxes. The efficiency at maximum power of each particle system decreases monotonously from an upper bound $\eta^+$ to a lower bound $\eta^-$ when $\Gamma$ increases from 0 to $\infty$, and leading to a region. It is recovered that the $\eta^+$ values for all three systems are bounded by $\eta_C/2 \leq \eta^+ \leq \eta_C/(2-\eta_C)$ due to strong coupling, where $\eta_C$ is the Carnot efficiency. For the Bose-Einstein system, it is found that the upper bound is approximated by the Curzon-Ahlborn efficiency: $\eta_{CA} = 1-\sqrt{1-\eta_C}$. When $\Gamma \to \infty$, the intrinsic maximum powers are proportional to the square of the temperature difference of two reservoirs for all three systems, and the corresponding lower bounds of efficiency at maximum power can be simplified in the same form of $\eta^- = \eta_C/[1+a_0(2-\eta_C)]$.

**PACS** number(s): 05.70.Ln, 05.30.-d

## 1. Introduction

The conversion of energy between thermal and other useful energy sources has attracted great interest in this energy-shortage world, and the possibility to manipulate this conversion seems of great significance. Many strategies have been proposed to find the suitable engine in various practical surroundings, including the construction of heat engines, the choice of the working substance, and some optimization measures on working conditions [1-13]. Among the possible energy conversion apparatuses, thermochemical engines are the easiest to control as they can work without moving parts [7-9,14-19]. Especially for thermoelectric engines, two-reservoir [8,16,17], three-reservoir [18,19], and four-reservoir [20,21] constructions have been proposed to produce electrical power from thermal energy effectively and in controllable manners. Focusing on the simple two-reservoir thermochemical engines, heat coupled particle flux can result in useful work between two different reservoirs (i.e., a hot one with temperature $T_h$ and a cold one with lower temperature $T_c$), adjusted by the chemical potentials gradient, the temperature gradient, and the transmission of particles [16,17].

It is well known that the efficiencies of all kinds of heat engines are bounded by the Carnot value $\eta_C = 1 - T_c/T_h$, which was originally obtained for a reversible Carnot heat engine [1]. It should be noted that the output power tends to zero for the reversible Carnot heat engine due to the infinite time of the working cycle. In practice, studying the maximum power has attracted much more attention, and Curzon and Ahlborn [10] derived an upper bound of efficiency at maximum power (EMP) from an endoreversible Carnot heat engine by using the finite-time technique, i.e., the so-called CA efficiency: $\eta_{CA} = 1 - \sqrt{1-\eta_C}$. In this kind of optimization, the irreversibility was accounted for through the finite time or the entropy production $\Delta S$, where $\Delta S > 0$ for any irreversible system based on the Second Law. By rearranging the entropy production, Esposito et al. [22] found that the EMP of low-dissipation Carnot engines $\eta^*$ are bounded by the condition $\eta_C/2 \leq \eta^* \leq \eta_C/(2-\eta_C)$, and the CA efficiency was recovered in the condition of symmetric dissipation. This region was supported by experimental data [23,24] and by other theoretical work on different models [25-27]. For thermochemical engines, the time dependence seems trivial due to the steady working state. However, this bounded result can still be recovered under the condition of strong coupling between the particle and heat fluxes [28]. With the weakening of the coupling, we will demonstrate in this paper that the EMP values might fall below the region introduced above.

*[*]tqiu@seu.edu.cn

To break the strong coupling, we introduce an energy filter with finite energy width to manipulate the transmission of particles between two reservoirs. The limit case of the filter with infinitesimal width corresponds to the strong coupling situation. Within reservoirs, near-independent particles, including Maxwell-Boltzmann (MB), Fermi-Dirac (FD) Bose-Einstein (BE) particles, are respectively considered as the working substance. Since some results about EMPs of heat engines with MB (e.g., the Feynman's ratchet [29] and the two-level atomic system [30]) and FD (almost all the electronic systems [31,32]) systems have already been reported, we here mainly focus on the engine with bosons in BE system, and evaluate the impact of the weak coupling (caused by the filter width) on the EMP.

## 2. Modeling the thermochemical engines

The thermochemical engines introduced here are actually exchanged-type energy conversion machines, constituted by two reservoirs and an energy filter between them, as shown briefly in Fig. 1(a). Filled with near-independent particles, the two reservoirs can be distinguished by their temperatures ($T_{c/h}$) and chemical potentials ($\mu_{c/h}$). In this two-reservoir system, particles with energy ($\varepsilon$) around the chemical potential may escape from one reservoir and flow into the other one due to the temperature and chemical potential gradients. The exchanged-type mechanism can be formed when the hot reservoir has a lower chemical potential with respect to the cold one (i.e., $\mu_c > \mu_h$). In this condition, some high-energy particles in the hot reservoir tend to diffuse to the cold reservoir due to the temperature gradient, and some low-energy particles in the cold reservoir diffuse to the hot one owing to the chemical potential gradient. In addition, the energy filter (visualized by a "demon" in Fig. 1(a)) between the two reservoirs can select the diffused particles based on their energy resulting in a net particle/heat flux, with adjustable direction and amount.

Generally, an energy filter can easily be realized by energy confinement effect in some tiny structures, e.g., quantum dot [33], quantum well [34], and other layered or periodic potential structures [35-37]. The transmission of particles through these energy filters can easily behave as a resonance with full width at half maximum, e.g. the Lorentz resonance. The center energy and full width at half maximum of those resonance profiles can be adjusted as wanted. For simplicity, the transmission probability of particles through the filter in this paper is chosen as a rectangular-type function [16] given by

$$\xi(\varepsilon) = \begin{cases} 1 & \varepsilon' < \varepsilon < \varepsilon' + \Delta\varepsilon \\ 0 & \text{elsewhere} \end{cases}, \quad (1)$$

where $\varepsilon'$ and $\Delta\varepsilon$ are the starting energy and energy width of the filter, respectively.

The two reservoirs considered here are infinite and invariable for fixed values of their temperatures and chemical potentials. The near-independent particles in the reservoirs can be described by the corresponding statistical distribution functions, i.e., the MB, FD, and BE distributions which can be expressed as $f_{MB} = e^{-r}$, $f_{BE} = 1/(e^r - 1)$, and $f_{FD} = 1/(e^r + 1)$, respectively, where $r = (\varepsilon - \mu)/k_B T$ and $k_B$ is the Boltzmann constant. It is assumed that the mean free path of the near-independent particles is larger than the size of the filter, and the interaction among particles is neglected inside the energy filter. Therefore, according to the Boltzmann equation [38], the particle flux flowing out of the hot/cold reservoir per unit time can be expressed as $\dot{N}_{h/c} = \pm C \int_0^\infty (f_h - f_c) \xi(\varepsilon) d\varepsilon$, where $C$ is a constant describing the diffusion rate of the particles. It should be noted that the diffusion rates of MB, FD, and BE particles are different from one another. For FD particles, this flux can actually be expressed by Landauer formula [8]. The reservoir absorbs/releases an average amount of heat $\varepsilon - \mu$ when one particle arrives/leaves, then the heat flux released by the hot/cold reservoir, per unit time, can be obtained by the integral equation:

$$\dot{Q}_{h/c} = \pm C \int_0^\infty (\varepsilon - \mu_{h/c})(f_h - f_c) \xi(\varepsilon) d\varepsilon, \quad (2)$$

where the $\pm$ signs refer to $h$ and $c$ respectively. When the energy filter selects particles with a single energy value $\varepsilon$, the heat flux will be strongly coupled with the particle flux since $\dot{Q}_{h/c} = (\varepsilon - \mu_{h/c}) \dot{N}_{h/c}$. However, this strong coupling will be broken if the width of the energy filter is finite.

Flowing against the chemical potential gradient, the net particle flux generates an output power $P = \dot{Q}_h + \dot{Q}_c$ and the efficiency of the engine can be calculated by $\eta = P/\dot{Q}_h$. From the distribution functions (also shown in Fig. 1(b)) and Eq. (2), it is found that the transmitted particles with energy $\varepsilon < \varepsilon_0 \equiv (\mu_c T_h - \mu_h T_c)/(T_h - T_c)$ induce $P < 0$ leading to a partial blocking of the engine, where $\varepsilon_0$ is the energy value obtained from $f_h - f_c = 0$. Therefore, particles with energy $\varepsilon \geq \varepsilon_0$ are strongly contributing to the thermochemical engine, pointing at a high significance of the energy filter in thermochemical engines.

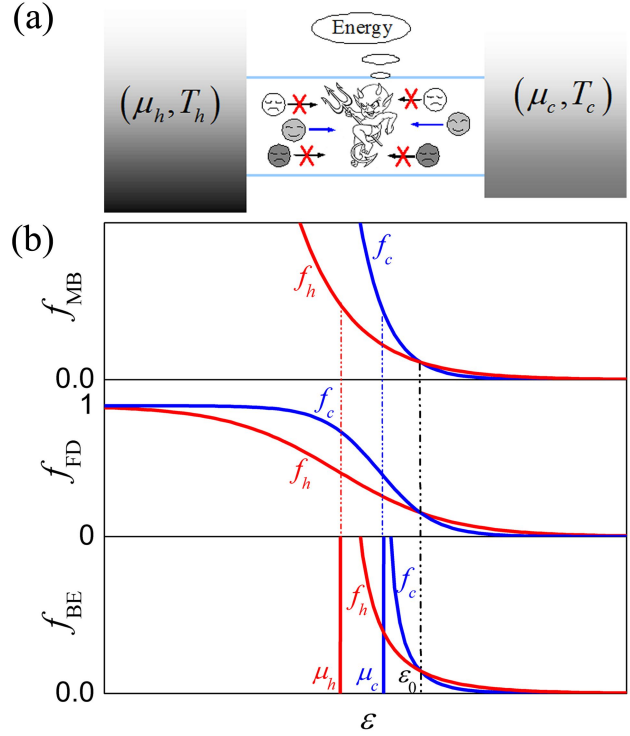

FIG. 1. (Color online) (a) Scheme of a thermochemical engine. A hot and a cold reservoirs (with chemical potential $\mu_{h/c}$ and temperature $T_{h/c}$, respectively) are connected to each other by an energy filter (the "demon" visualized in the picture), which blocks some diffused particles depending on their energy (see the red cross). Particles successfully filtered upon their energy can be exchanged between the two reservoirs. (b) Schematic of distribution functions with respect to particle energy $\varepsilon$ for MB, FD, and BE systems when $\mu_h < \mu_c$.

Substitution of Eq. (1) into Eq. (2) yields the fundamental equation for this work, $\dot{Q}_{h/c} = \pm C\int_{\varepsilon'}^{\varepsilon'+\Delta\varepsilon}(\varepsilon - \mu_{h/c})(f_h - f_c)d\varepsilon$, which can be finally expressed as a function of temperature $T_h$, dimensionless scaled energies $r_{h/c} = (\varepsilon' - \mu_{h/c})/k_B T_{h/c}$, width $\Gamma = \Delta\varepsilon/[k_B(T_c + T_h)/2]$, and Carnot efficiency $\eta_C$. Thus, the output power of the engine can be expressed as the function of

$$P = Ck_B^2 T_h^2 F(r_c, r_h, \Gamma, \eta_C). \qquad (3)$$

It should be noted that the temperature considered here is not infinite, so that $\Gamma$ is acceptable to stand for the energy width of the filter $\Delta\varepsilon$. From Eq. (3), it is clearly shown that the maximum power can be obtained by $\partial_{r_h} P = \partial_{r_c} P = 0$ at given values of $T_h$, $\Gamma$, and $\eta_C$. Actually, $T_h$ does not directly influence the efficiency at maximum power, while it affects the maximum power of the engine. For convenience, $T_h$ will be fixed at an arbitrary value in the following calculation, so that $P^* = P/Ck_B^2 T_h^2$ can represent the engine power.

## 3. Optimum analysis at maximum power for BE system

From the analysis above, the expressions of heat flux, power, and efficiency of thermochemical engines can be expressed for MB, FD, and BE systems. Here we focus on the BE system (others are shown in Appendix A), whose parameters are given by the following equations:

$$P^* = \left[r_h - (1-\eta_C)r_c\right]\left[\ln\left(\frac{e^{r_h} - e^{-\Gamma_h}}{e^{r_h}-1}\right) - (1-\eta_C)\ln\left(\frac{e^{r_c}-e^{-\Gamma_c}}{e^{r_c}-1}\right)\right] \quad (4)$$

$$\dot{Q}_h^* = \left\{\text{Li}_2\left(1-e^{-r_h-\Gamma_h}\right) - \text{Li}_2\left(1-e^{-r_h}\right) - (1-\eta_C)^2\left[\text{Li}_2\left(1-e^{-r_c-\Gamma_c}\right) - \text{Li}_2\left(1-e^{-r_c}\right)\right]\right.$$
$$\left. - (1-\eta_C)\left[r_h - (1-\eta_C)r_c\right]\left[\ln\left(1-e^{-r_c-\Gamma_c}\right) - \ln\left(1-e^{-r_c}\right)\right]\right\} \quad (5)$$

where $\Gamma_h = \Delta\varepsilon/k_B T_h = \Gamma(1-\eta_C/2)$, $\Gamma_c = \Delta\varepsilon/k_B T_c = \Gamma(1-\eta_C/2)/(1-\eta_C)$, $\text{Li}_n(x) = \sum_{i=1}^{\infty}\frac{x^i}{i^n}$ is the polylogarithm function, and $\dot{Q}_h^* = \dot{Q}_h / Ck_B^2 T_h^2$. Therefore, the temperature-independent EMP can be obtained for different filter widths $\Gamma$, although its derivation can only occur through numerical calculations.

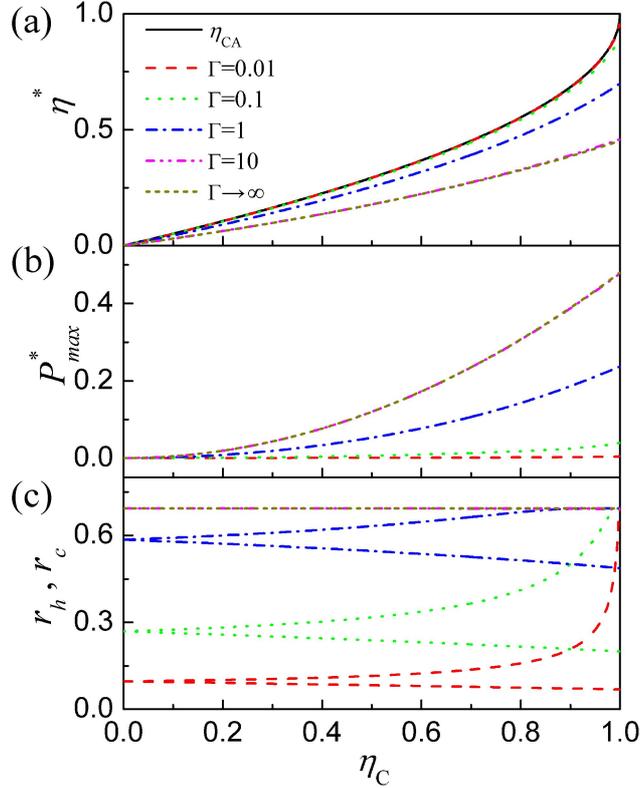

FIG. 2. (Color online) (a) Efficiency at maximum power of a thermochemical engine with BE system for different filter widths. The solid line indicates the CA efficiency, which seems to act as an upper bound for EMP when $\Gamma \to 0$; the lower bound can be obtained when $\Gamma \to \infty$. (b) Maximum power for different filter widths. (c) Each pair of curves denotes the corresponding dimensionless scaled energies $r_{h/c}$ ($r_c \geq r_h$), which seem to decrease from about 0.693147 to 0 when $\Gamma$ narrows down from $\infty$ to 0.

Fig. 2. shows the numerically calculated results for the EMP of a thermochemical engine with a BE system. The EMP $\eta^*$ increases with the Carnot efficiency $\eta_C$ at a fixed filter width $\Gamma$.

Smaller $\Gamma$ values yield higher $\eta^*$, while gradually causing the output power to vanish. From Fig. 2(a), it is possible to see that $\eta^*$ increases from a definite lower bound $\eta^-$ to a seemingly upper bound $\eta^+$ at a fixed $\eta_C$ when $\Gamma$ decreases from $\infty$ to 0 leading to the formation of an EMP region. Accordingly, the dimensionless scaled energies $r_{c/h}$ decrease from about 0.693147 to 0 at an arbitrary $\eta_C$. It is also found that $\eta^*$ is very close to a lower bound even if $\Gamma=10$, which happens because particle transmission only takes place for particles whose energy differs from the chemical potential by few $k_B T$. Moreover, the starting energy of the filter is always $\varepsilon' \geq \varepsilon_0$ due to the numerical result of $r_c \geq r_h$ (as shown in Fig. 2(c)), which indicates that no particle blocks the engine under the condition of maximum power.

To study the EMP region further, the two limit cases of $\Gamma \to 0$ and $\Gamma \to \infty$ are discussed in depth as follows (also focusing on a BE system, while others are shown in Appendix B), where the upper and lower bounds may be obtained. When $\Gamma \to 0$, the heat flux out of reservoirs can be approximated as

$$\dot{Q}_{h/c} \approx \pm C k_B T_{h/c} r_{h/c} \left( \frac{1}{e^{r_h}-1} - \frac{1}{e^{r_c}-1} \right) \delta\varepsilon, \tag{6}$$

where $\delta\varepsilon$ indicates the infinitesimal energy change around $\varepsilon$, and the heat flux is strongly coupled with the particle flux due to the equation $\dot{Q}_{h/c} = (\varepsilon' - \mu_{h/c}) \dot{N}_{h/c}$. It should be noted that the energy of transmitted particles is just the starting energy of the filter in this situation, i.e., $\varepsilon = \varepsilon'$. Combining with the power $P = \dot{Q}_h + \dot{Q}_c$, the efficiency can be expressed as $\eta = 1-(1-\eta_C) r_c / r_h$.

By imposing $\partial_{r_h} P = \partial_{r_c} P = 0$, one can find the relationship $\frac{\sinh(r_h/2)}{\sinh(r_c/2)} = \sqrt{1-\eta_C}$ at maximum power. Unfortunately, no analytical solution or even numerical result (due to the limited numerical capability) is obtained except that of $r_{c/h} \to 0$. In this case the EMP is expressed as

$$\eta^* = 1 - 2(1-\eta_C) \frac{\operatorname{arcsinh}\left[\sinh(r_h/2)/\sqrt{1-\eta_C}\right]}{r_h}, \tag{7}$$

and it is found that the zero order term of this expression is just the CA efficiency:

$$\eta^* = 1 - \sqrt{1-\eta_C} + O(r_h^2). \tag{8}$$

Since $r_h \to 0$ when $\Gamma \to 0$, we strongly suggest that the upper bound of the EMP $\eta^+$ region can be approximated as $\eta_{CA}$.

When $\Gamma \to \infty$, it can be found that $r_c = r_h \approx 0.693147$ from the numerical results of Fig. 2. One can then deduce the starting energy of the filter $\varepsilon' = \varepsilon_0$ from $r_c = r_h$, which means that all transmitted particles sustain the good working condition of the heat engine since their energy is located in the working region $\varepsilon \geq \varepsilon_0$. In fact, the power can be initially maximized with respect to the starting energy $\varepsilon'$ of the filter, and can be expressed as:

$$P' = -Ck_B^2 T_h^2 \eta_C^2 r_0 \ln\left(1 - e^{-r_0}\right) \tag{9}$$

where $r_0 \equiv \dfrac{\mu_c - \mu_h}{k_B(T_h - T_c)}$ and the value of $r_0$ is determined by $\partial P' / \partial r_0 = 0$ for the maximum power $P_{\max}$. Then the EMP for the width $\Gamma \to \infty$, i.e., the lower bound of the EMP region, is

$$\eta^- = \frac{\eta_C}{1 + a_0(2 - \eta_C)}, \tag{10}$$

where $a_0 \approx 1.21186$. We can note that in this case $P_{\max}$ indicates the intrinsic maximum power that this thermochemical engine can achieve, and $P_{\max} \propto \Delta T^2$ where $\Delta T = T_h - T_c$, which has also been found for quantum thermoelectric heat engines [32].

## 4. EMP regions for MB and FD systems

Following a similar approach, the EMP regions for MB and FD systems can also be obtained, as calculated in Appendix B. The upper bounds obtained in references [29,30] and [31,32] are recovered here for MB and FD systems, respectively. Focusing on these upper bounds $\eta^+$, it is clearly found from Fig. 3 that $\eta_C/2 \leq \eta^+ \leq \eta_C/(2 - \eta_C)$ for all of MB, FD, and BE systems. Therefore, our results are consistent with reference [22] because these upper bounds are obtained under the condition of strong coupling between the particle and heat fluxes. Furthermore, one can find that $\eta_{FD}^+ > \eta_{MB}^+ > \eta_{BE}^+ \approx \eta_{CA}$ at an arbitrary given Carnot efficiency, although the values are very close to each other. As $\Gamma$ increases, the coupling is weakened leading to a lower EMP and even $\eta^* < \eta_C/2$ when $\Gamma$ is large enough. For $\Gamma \to \infty$, the lower bounds of the EMP for MB and FD systems can be calculated in the same way as for BE systems, obtaining the same form of Eq. (10), with the parameters shown in Table I. The same inequality $\eta_{FD}^- > \eta_{MB}^- > \eta_{BE}^-$ is also found for the lower bounds.

TABLE I. Numerical parameters $a_0$ of the lower bound $\eta^- = \eta_C / [1 + a_0(2 - \eta_C)]$ of EMP for MB, FD, and BE systems when $\Gamma \to \infty$.

|       | MB | FD      | BE      |
|-------|----|---------|---------|
| $a_0$ | 1  | 0.93593 | 1.21186 |

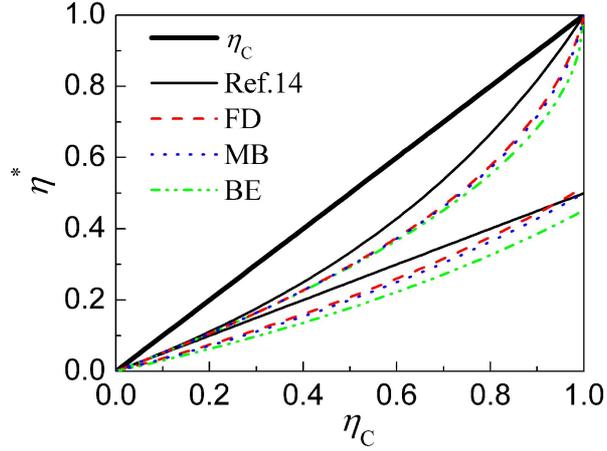

FIG. 3. (Color online) Upper and lower bounds of the EMP regions for MB, FD, and BE systems. Two thin solid lines depict the EMP region under the condition of strong coupling [22], which contains all the upper bounds for MB, FD, and BE systems.

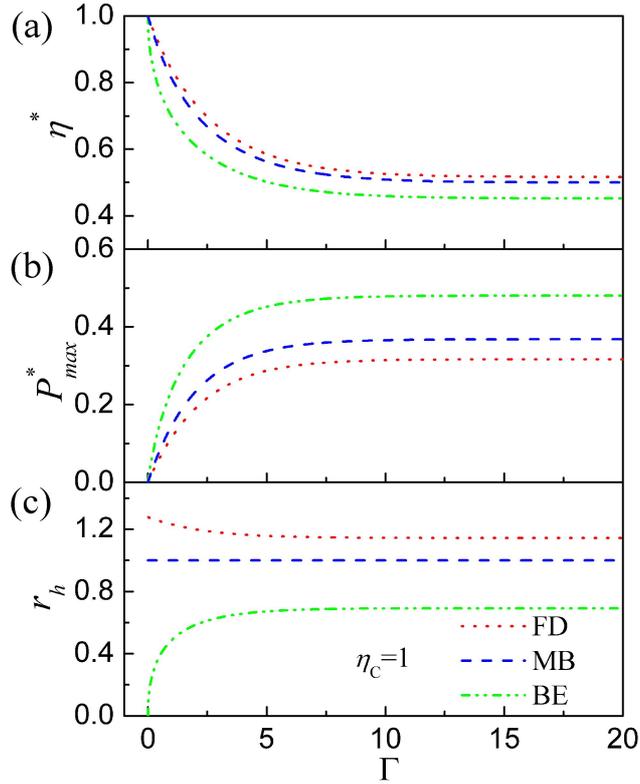

FIG. 4. (Color online) (a) EMPs, (b) maximum powers, and (c) the corresponding $r_h$, as a function of the filter width for MB, FD, and BE systems, when the Carnot efficiency $\eta_C = 1$.

In order to investigate the relationship between the EMP $\eta^*$ and the filter width $\Gamma$, we calculated several sets of $\eta^*$ at different $\eta_C$ values. As $\Gamma$ increases gradually starting from 0, the EMPs decrease monotonously from the upper bound $\eta^+$ and the maximum powers increase monotonously from 0 for MB, FD, and BE systems. Both the EMP and the maximum power tend to saturate when $\Gamma$ increases to about 10. We here only show the situation of $\eta_C = 1$ in Fig. 4. It is clearly shown that only the particles whose energy differs from chemical potential by few $k_B T$ contribute to our thermochemical engines. At a fixed width of the filter, the EMPs of different systems satisfy the inequality $\eta_{FD}^* > \eta_{MB}^* > \eta_{BE}^*$ at a given $\eta_C$. Additionally, $r_h \to 0$ when $\Gamma \to 0$ is obtained again for BE systems.

## 6. Conclusions

Two-reservoir models of thermochemical engines with MB, FD, and BE particles as a working substance were studied respectively. With the help of the temperature gradient and the chemical potential gradient, particles can diffuse between the two reservoirs. Simultaneously, the heat flux coupled with particle flux will be transferred or converted into useful work. After introducing a rectangular-type energy filter with width $\Gamma$, the power and the efficiency of the engine can be described by simplified expressions directly for MB, FD, and BE systems, respectively. Based on these expressions, numerical or analytical solutions of temperature-independent EMPs are obtained at a given width $\Gamma$, finding that an EMP region is formed for each system when $\Gamma$ increases from 0 to $\infty$. The upper $\eta^+$ and lower bound $\eta^-$ of the EMP region are obtained at the limits of $\Gamma \to 0$ and $\Gamma \to \infty$ respectively. When $\Gamma \to 0$, the heat flux of the thermochemical engine is strongly coupled with the particle flux, and the upper bounds for the three systems are bounded by $\eta_C / 2 \leq \eta^+ \leq \eta_C / (2 - \eta_C)$, which is consistent with the results in reference [22]. Focusing on the BE system, the upper bound of EMP could be just approximated by the CA efficiency. As the filter width increases, the coupling is weakened, and the EMP can reach values lower than $\eta_C / 2$ when $\Gamma$ is large enough. When $\Gamma \to \infty$, the intrinsic maximum power of the thermochemical engines is obtained for all of MB, FD, and BE systems, finding that the maximum power $P_{max} \propto \Delta T^2$ in this case. Moreover, the lower bounds of the EMP for all systems can be simplified in the same form of $\eta^- = \eta_C / [1 + a_0 (2 - \eta_C)]$. It should be noted that the heat loss of the engines was neglected in all the analysis above, so that our results correspond to the theoretical limits which are more favorable than what can be obtained in practice.

## Acknowledgments

This work was supported by the Fundamental Research Funds for the Central Universities and Research and Innovation Project for College Graduates of Jiangsu Province (No. CXZZ13_0081), People's Republic of China.

## Appendix A: Heat flux of thermochemical engines for MB and FD system

From Eqs. (1), (2), and (3), the power of a thermochemical engine with a MB system can be expressed as

$$P^* = \left[ r_h - (1 - \eta_C) r_c \right] \left[ -e^{-r_h - \Gamma_h} + e^{-r_h} + (1 - \eta_C)\left( e^{-r_c - \Gamma_c} - e^{-r_c} \right) \right], \quad (A1)$$

and the heat flux out of the hot reservoir is given by

$$\dot{Q}_h^* = \{-(1+r_h+\Gamma_h)e^{-r_h-\Gamma_h} + (1+r_h)e^{-r_h} + (1-\eta_C)^2\left[(1+r_c+\Gamma_c)e^{-r_c-\Gamma_c} - (1+r_c)e^{-r_c}\right]$$

$$+(1-\eta_C)\left[r_h - (1-\eta_C)r_c\right]\left(e^{-r_c-\Gamma_c} - e^{-r_c}\right)\}. \tag{A2}$$

For a FD system [31], the power and heat fluxes out of the hot reservoir are:

$$P^* = \left[r_h - (1-\eta_C)r_c\right]\left[(1-\eta_C)\ln\left(\frac{1+e^{-r_c-\Gamma_c}}{1+e^{-r_c}}\right) - \ln\left(\frac{1+e^{-r_h-\Gamma_h}}{1+e^{-r_h}}\right)\right] \tag{A3}$$

and

$$\dot{Q}_h^* = \{\left[g(e^{-r_h-\Gamma_h}) - g(e^{-r_h}) + \text{Li}_2(-e^{-r_h-\Gamma_h}) - \text{Li}_2(-e^{-r_h})\right] - (1-\eta_C)^2\left[g(e^{-r_c-\Gamma_c}) - g(e^{-r_c})\right]$$

$$+\text{Li}_2(-e^{-r_c-\Gamma_c}) - \text{Li}_2(-e^{-r_c})\right] + \left[(1-\eta_C)r_h - (1-\eta_C)^2 r_c\right]\cdot\left[\ln(1+e^{-r_c-\Gamma_c}) - \ln(1+e^{-r_c})\right]\}, \tag{A4}$$

respectively, where $g(x) = \ln(x)\ln(1+x)$.

## Appendix B: Maximum power at $\Gamma \to 0$ and $\Gamma \to \infty$

When the filter width vanishes, i.e., $\Gamma \to 0$, the power and heat fluxes out of hot reservoir can be expressed as

$$P \approx Ck_B T_h \left[r_h - (1-\eta_C)r_c\right](f_h - f_c)\delta\varepsilon \tag{B1}$$

and

$$\dot{Q}_h \approx Ck_B T_h r_h (f_h - f_c)\delta\varepsilon \tag{B2}$$

where $f_{h/c} = \frac{1}{e^{r_{h/c}} - 1}$, $f_{h/c} = e^{-r_{h/c}}$, and $f_{h/c} = \frac{1}{e^{r_{h/c}} + 1}$ for BE, MB, and FD systems, respectively.

Their efficiencies can thus be written in the same form, i.e. $\eta = 1 - (1-\eta_C)r_c/r_h$. It is possible to calculate the EMP from $\partial_{r_h} P = \partial_{r_c} P = 0$. It can be found that the EMP for a MB system [29,30] has the analytical expression $\eta^* = \eta_C^2 / \left[\eta_C - (1-\eta_C)\ln(1-\eta_C)\right]$, and the EMP for BE or FD [31,32] systems can only be obtained by numerical calculation.

When $\Gamma \to \infty$, the power can be initially maximized with respect to the starting energy $\varepsilon'$ of the filter, i.e. $P' = C\int_{\varepsilon_0}^{\infty}(\mu_c - \mu_h)(f_h - f_c)d\varepsilon$, which can be expressed analytically as:

$$P' = \begin{cases} Ck_B^2 T_h^2 \eta_C^2 r_0 e^{-r_0} \\ Ck_B^2 T_h^2 \eta_C^2 r_0 \ln(1+e^{-r_0}) \\ -Ck_B^2 T_h^2 \eta_C^2 r_0 \ln(1-e^{-r_0}) \end{cases} \tag{B3}$$

for MB, FD, and BE in turn. The corresponding heat flux out of hot reservoir is

$$\dot{Q}'_h = \begin{cases} Ck_B^2 T_h^2 \eta_C (2 + r_0 - \eta_C) e^{-r_0} \\ Ck_B^2 T_h^2 \eta_C \left[ r_0 \ln(1 + e^{-r_0}) + (\eta_C - 2) \text{Li}_2(-e^{-r_0}) \right] \\ Ck_B^2 T_h^2 \eta_C \left\{ (2 - \eta_C) \left[ \frac{\pi^2}{6} - \text{Li}_2(1 - e^{-r_0}) \right] + (1 - \eta_C) r_0 \ln(1 - e^{-r_0}) \right\} \end{cases} \quad (B4)$$

where $r_0 \equiv \frac{\mu_c - \mu_h}{k_B(T_h - T_c)}$ for all the cases. It should be noted that $P'$ is not the maximum power we want because it is still dependent on chemical potentials. After further optimization, maximum power $P_{max}$ can be obtained through $\partial P'/\partial r_0 = 0$, and the values of $r_0 \approx 1$, 1.14455, and 0.693147 can also be determined at maximum power for MB, FD, and BE systems, respectively. The EMP can then be obtained by $\eta^* = P_{max}/\dot{Q}_h \big|_{\max P}$. Focusing on the maximum power in all the cases when $\Gamma \to \infty$, it is found that $P_{max} \propto \Delta T^2$ due to the equation $T_h \eta_C = T_h - T_c = \Delta T$.

## References

[1] S. Carnot. *Réflexions sur la Puissance Motrice du Feu et sur les Machines Propres à Développer cette Puissance* (Bachelier, Paris, 1824).

[2] U. Seifert, Rep. Prog. Phys. **75**, 126001 (2012).

[3] E. Geva and R. Kosloff, J. Chem. Phys. **96**, 3054 (1992).

[4] H. T. Quan, Y. Liu, C. P. Sun, and F. Nori, Phys. Rev. E **76**, 031105 (2007).

[5] J. Wang, Z. Ye, Y. Lai, W. Li, and J. He, Phys. Rev. E **91**, 062134 (2015).

[6] J. Guo, J. Wang, Y. Wang, and J. Chen, Phys. Rev. E **87**, 012133 (2013).

[7] B. Sothmann, R. Sánchez, A. N. Jordan, and M. Büttiker, Phys. Rev. B **85**, 205301 (2012).

[8] T. E. Humphrey, R. Newbury, R. P. Taylor, and H. Linke, Phys. Rev. Lett. **89**, 116801 (2002).

[9] Y. Zhang, G. Lin, J. Chen, Phys. Rev. E **91**, 052118 (2015).

[10] F. L. Curzon and B. Ahlborn, Am. J. Phys. **43**, 22 (1975).

[11] T. Schmiedl and U. Seifert, EPL **81**, 20003 (2008).

[12] Z. Tu, Chinese Phys. B **21**, 020513 (2012).

[13] N. Sánchez-Salas, L. López-Palacios, S. Velasco, and A. C. Hernández, Phys. Rev. E **82**, 051101 (2010).

[14] B. Lin and J. Chen, J. Phys. A: Math. Theor. **42**, 075006 (2009).

[15] X. Luo, N. Liu, and J. He, Phys. Rev. E **87**, 022139 (2013).

[16] M. F. O'Dwyer, R. A. Lewis, C. Zhang, and T. E. Humphrey, Phys. Rev. B **72**, 205330 (2005).

[17] M. Esposito, K. Lindenberg, and C. Van den Broeck, EPL **85**, 60010 (2009).

[18] R. Sánchez and M. Büttiker, Phys. Rev. B **83**, 085428 (2011).

[19] O. Entin-Wohlman, Y. Imry, and A. Aharony, Phys. Rev. B **82**, 115314 (2010)

[20] J. Stark, K. Brandner, K. Saito, and U. Seifert, Phys. Rev. Lett. **112**, 140601 (2014).

[21] B. Sothmann, R. Sánchez, and A. N. Jordan, EPL **107**, 47003 (2014)

[22] M. Esposito, R. Kawai, K. Lindenberg, and C. Van den Broeck, Phys. Rev. Lett. **105**, 150603 (2010).

[23] H. B. Callen, *Thermodynamics and an Introduction to Thermostatistics* (Wiley, New York, 1985), 2nd ed..

[24] A. Bejan, *Advanced Engineering Thermodynamics* (Wiley, New York, 1997), p. 377.

[25] C. Van den Broeck, N. Kumar, and K. Lindenberg, Phys. Rev. Lett. **21**, 210602 (2012).

[26] Y. Wang and Z. Tu, Phys. Rev. E **85**, 011127 (2012).

[27] J. Wang, J. He, and Z. Wu, Phys. Rev. E **85**, 031145 (2012); J. Wang and J. He, Phys. Rev. E **86**, 051112 (2012).

[28] M. Esposito, K. Lindenberg, and C. Van den Broeck, Phys. Rev. Lett. **102**, 130602 (2009).

[29] Z. Tu, J. Phys. A: Math. Theor. **41**, 312003 (2008).

[30] R. Wang, J. Wang, J. He, and Y. Ma, Phys. Rev. E **87**, 042119 (2013).

[31] X. Luo, C. Li, N. Liu, R. Li, J. He, and T. Qiu, Phys. Lett. A **377**, 1566 (2013); X. Luo, J. He, K. Long, J. Wang, N. Liu, and T. Qiu, J. Appl. Phys. **115**, 244306 (2014).

[32] R. S. Whitney, Phys. Rev. Lett. **112**, 130601 (2014); R. S. Whitney, Phys. Rev. B **91**, 115425 (2015).

[33] B. Sothmann, R. Sánchez, and A. N. Jordan, Nanotechnology **26**, 032001 (2015).

[34] N. Liu, X. Luo, and M. Zhang, Chinese Phys. B **23**, 080502 (2014); X. Luo, N. Liu, J. He, and T. Qiu, Appl. Phys. A **117**, 1031 (2014).

[35] P. Hänggi and F. Marchesoni, Rev. Mod. Phys. **81**, 387(2009).

[36] L. P. Faucheux, L. S. Bourdieu, P. D. Kaplan, and A. J. Libchaber, Phys. Rev. Lett. 74, 1504 (1995).

[37] M. Barbier, F. M. Peeters, P. Vasilopoulos, and J. Milto Pereira Jr., Phys. Rev. B 77, 115446 (2008).

[38] G. M. Kremer, *An Introduction to the Boltzmann Equation and Transport Processes in Gases* (Springer, Verlag Berlin Heidelberg, 2010), p.37-42.